\documentclass[11pt]{article}
\usepackage{amsmath}
\usepackage{amssymb}
\usepackage{latexsym}
\usepackage{epsfig}

\newcommand{\be}{\begin{equation}}
\newcommand{\ee}{\end{equation}}

\newcommand{\sectiono}[1]{\section{#1}\setcounter{equation}{0}}

\newcommand{\p}{\partial}

\begin{document}

{}~
\hfill\vbox{\hbox{hep-th/0506076}\hbox{MIT-CTP-3648}
}\break

\vskip 3.0cm

\centerline{\Large \bf Rolling Closed String Tachyons and the Big Crunch}

\vspace*{10.0ex}

\centerline{\large Haitang Yang
and Barton Zwiebach}

\vspace*{7.0ex}

\vspace*{4.0ex}

\centerline{\large \it  Center for Theoretical Physics}

\centerline{\large \it
Massachusetts Institute of Technology}

\centerline{\large \it Cambridge,
MA 02139, USA}
\vspace*{1.0ex}

\centerline{hyanga@mit.edu, zwiebach@lns.mit.edu}

\vspace*{10.0ex}

\centerline{\bf Abstract}
\bigskip
\smallskip

We study the low-energy effective field equations that couple
gravity, the dilaton, and the bulk closed
string tachyon of bosonic closed string theory.
We establish that whenever the tachyon induces the rolling
process, the string metric remains fixed while the dilaton
rolls to strong coupling. For negative definite potentials
we show that this results in an Einstein metric that
crunches the universe in finite time. This behavior is shown
to be rather generic even if the potentials are not negative
definite.  The  solutions are reminiscent
of those in the collapse stage of a cyclic universe cosmology
where scalar field potentials with negative energies play a central role.

\vfill \eject

\baselineskip=16pt

\vspace*{10.0ex}

%\tableofcontents

\sectiono{Introduction}

A tachyon of closed string theory is said to be
a bulk tachyon if it lives throughout spacetime.
In the presence of a tachyon one has an instability
and two important and related questions arise:
\begin{enumerate}
\item
Is there a ground state of the theory without the instability ?
\item   What is the end-result
of the physical decay process associated with the instability ?
\end{enumerate}

The answers are presently known for the open string theory
tachyons that live on the world-volume of
unstable D-branes~\cite{reviews}. The ground state, or
tachyon vacuum, is a state without the D-brane and
without open strings -- it is in fact
the vacuum state of closed strings.
In the associated physical decay process
the D-brane dissappears but the result is not quite the
closed string vacuum but rather an excited state of closed
strings that carries the original energy of the D-brane.
The decay process is not simply a transition from the unstable
to the stable vacuum.

The purpose of the present paper is to study
 the physical decay induced by the bulk closed string tachyon
of bosonic closed string theory -- the second
question above applied to bulk tachyons (for localized
closed string tachyons, see~\cite{Headrick:2004hz}.).
Our work was prompted by
new information about the first question:
recently-found evidence that the
closed string field theory ``tachyon potential" has
a critical point -- a candidate for a closed string tachyon vacuum~\cite{HZ}.
A set of considerations suggests that in such vacuum
closed string states would not propagate and
spacetime would cease to be dynamical.  Our analysis of
the physical decay aims to illuminate the nature of the tachyon vacuum.
This may be possible because the physical decay turns out to be
rather insensitive to the specific details of the tachyon
potential, about which little is known.

We study here the low-energy field equations that couple
the metric, the dilaton, and the tachyon.  These equations are
motivated by the conditions of conformal invariance of
sigma models~\cite{sdas} and are expected to provide solutions
that capture relevant features of exact string theory solutions.
The low-energy field equations
have been used in many papers to study all kinds of
dynamical and cosmological issues (for a review and references
see~\cite{Gasperini:2002bn}). Few of these works, however, deal with
key features of our present problem: a standard minimal
coupling of the dilaton to other fields, an unstable closed
string vacuum with zero cosmological
constant, a tachyon potential that is not positive, and a
rolling process induced by the tachyon.
The related problem of light (bulk) tachyons that
arise from circle compactification has been studied by
Dine{\em \,et.al.}~\cite{Dine:2003ca} and
Suyama~\cite{Suyama:2003as}.
These authors  compute quadratic and quartic terms in the
tachyon potential and consider a cosmological evolution that
involves the metric, the dilaton, the radion, and the tachyon.
The authors of~\cite{Dine:2003ca} state that numerical studies
show that rather general initial
conditions lead to a radius that evolves to make the tachyon more
tachyonic and a dilaton that evolves to make the system strongly coupled
(the simplicity of a light tachyon seems to be illusory).
The author of~\cite{Suyama:2003as} freezes the radion and discusses
explicitly the simplified  system, showing in a numerical solution
how it appears to be driven to strong coupling.

Our analysis assumes {\em arbitrary} tachyonic
potentials $V(T) = -{1\over 2} m^2 T^2 + \mathcal{O}(T^3)$
and reveals a few surprises.  We have found
that if the rolling process is triggered by the tachyon  the
string metric does {\em not} evolve. Moreover, the dilaton expectation
value $\Phi$ will always increase as time goes by.
If the tachyon history $T(t)$ is such that the potential
is negative, $V(T(t)) \leq 0$, the evolution reaches a singular point in finite
time: both $\dot  T$ and $\dot \Phi$
become infinite, and so do $T$ and $\Phi$.  In the string
frame this is a  system with infinite string coupling, while
in the Einstein frame the universe undergoes a big crunch.
While negative potentials help accelerate its occurrance,
a crunch occurs at finite time (both in the string and Einstein
frames) for a wide class
of  potentials that grow arbitrarily large and positive for
large $T$,  $V(T)= - T^2 + T^4$,
for example. The growing dilaton acts on
the tachyon like anti-friction, a force proportional to the
tachyon velocity in the direction of the velocity. This generally
enables the tachyon to reach infinite value in finite time, even
if it has to climb an infinite potential.

This paper is organized as follows.  In Section 2 we write the relevant
coupled equations and examine them in the cosmological setting.  We use
the string metric and emphasize how the dilaton time derivative plays
a role similar to that of {\em minus} the Hubble parameter $H(t)$. In Section 3 we
define tachyon-induced rolling and show that it results in a
constant string metric. The rolling problem simplifies considerably and
becomes the coupled dynamics of a dilaton and a tachyon. Analytic solutions
are possible if one can solve a certain first-order nonlinear differential
equation. Up to numerical
constants, the dilaton-tachyon
equations can be mapped to those that describe a single scalar field
rolling in Einstein's theory.  In Section 4 we establish
that the Einstein
metric crunches in finite time if the tachyon potential
is negative throughout the rolling solution.  In Section 5 we consider potentials
that can be positive and develop tools to decide if there is a big crunch
and if it occurs in finite time. Conclusions are offered in
Section 6.

\sectiono{The coupled system of rolling fields}

Consider the  action  that describes
the low-energy dynamics of
the metric, the dilaton, and the tachyon:
 \begin{equation}
\label{sigma_action}
S=\frac{1}{2\kappa^2 }\int \, d^{\,d+1} x
\sqrt{- g}\, e^{-2\Phi}\Bigl(R+4 (\p_\mu\Phi)^2  -
(\p_\mu T)^2  -2 V(T)\,\Bigr)\,.
\end{equation}
Here $g_{\mu\nu}$ is the string metric, $\Phi$ is
the dilaton, and $T$ is the tachyon, with potential
$V(T)$.  The number of spatial
dimensions is $d$. We are following
the conventions of~\cite{Kostelecky:1992vg}, with their
dilaton $\phi$ replaced by $(-2\Phi)$. The metric--dilaton
part of the action is that in~\cite{Polchinski:1998rq}.
The equations of motion are:
\begin{equation}
\label{smeqmotion}
\begin{split}
R_{\mu\nu} +
2 \nabla_\mu\nabla_\nu \Phi  - (\p_\mu T) (\p_\nu T) &= 0\,, \\[0.5ex]
\nabla^2 T - 2 (\p_\mu \Phi)  (\p^\mu T) - V' (T) &= 0\,, \\[0.5ex]
\nabla^2 \Phi - 2 (\p_\mu \Phi)^2 - V(T) &= 0\,.
\end{split}
\end{equation}
To evaluate the action on-shell we multiply the first equation by $g^{\mu\nu}$,
use the third equation  to eliminate $\nabla^2 \Phi$, and find that
$R + 4 \,(\p_\mu \Phi)^2 - (\p_\mu T)^2 = - 2 V(T)$.
Using this,
 \begin{equation}
\label{sigma_action_os}
S_{on-shell}=\frac{1}{2\kappa^2 }\int \, d^{\,d+1} x
\sqrt{- g}\, e^{-2\Phi}\bigl( -4 V(T)\,\bigr)\,.
\end{equation}

We look for solutions of (\ref{smeqmotion}) that
represent a rolling tachyon field $T(t)$ accompanied
by a time dependent dilaton $\Phi(t)$ and a time
dependent string metric of the form
\begin{equation}
ds^2 = - (dt)^2 + a^2 (t)  \bigl( dx_1^2 + dx_2^2 +
\ldots  dx_{d}^2 \bigr) \,, \quad  H(t) \equiv {\dot a (t) \over a(t)} \,.
\end{equation}
With this metric, the gravitational equations of motion (first line
in (\ref{smeqmotion})) give two equations
\begin{equation}
\label{graveqn}
d\, {\ddot a\over a} + \dot T^2  - 2\, \ddot \Phi = 0\,,\quad
{\ddot a\over a} + (d-1) \Bigl({\dot a\over a}\Bigr)^2
- 2{\dot a\over a} \,\dot \Phi  =0 \,.
\end{equation}
The equations of motion for the dilaton and the tachyon are:
\begin{eqnarray}
\label{dilroll99}
\ddot \Phi + \bigl(\, d\, H - 2\,\dot \Phi \bigr) ~\dot \Phi  + V(T) &=& 0\,,\\[0.6ex]
\label{tachroll99}
\ddot T + \bigl(\, d\, H - 2\,\dot \Phi \bigr) \dot T  + V'(T) &=& 0\,.
\end{eqnarray}
We recognize  the familiar Hubble ``friction" term that couples
$H$ to the field velocity.  Indeed, for $H>0$
the force is opposite to the velocity and slows down the
field.  Similarly,
the dilaton velocity $\dot \Phi$ is {\em anti-}friction:  if $\dot \Phi >0$
the force is in the direction of the velocity and accelerates the field.
The dilaton is driven by $-V(T)$; it will tend to go to strong coupling
while  $V(T)<0$.

The gravity equations (\ref{graveqn}) can be rearranged into two equivalent
equations:
\begin{eqnarray}
\label{dotHeqn}
{\textstyle{1\over 2}}\, (d-1) \, \dot H &=& -{\textstyle{1\over 2}}
\,\dot T^2 + \, \ddot \Phi -
\, H \, \dot \Phi \,, \\[1.0ex]
\label{nodotHeqner}
{\textstyle{1\over 2}} \,d \,(d-1) \, H^2  &=&
 {\textstyle{1\over 2}} \,\dot T^2 -  \ddot \Phi + d\, H \, \dot \Phi  \,.
\end{eqnarray}
It can be shown that if equation (\ref{nodotHeqner}) holds at some time, equations
(\ref{tachroll99}), (\ref{dilroll99}), and (\ref{dotHeqn}) guarantee that
it holds for all times.

It is instructive to  compare of the previous equations
with those that govern the dynamics  of a scalar field
$\phi$ with potential $V(\phi)$ coupled to gravity {\em without}
a dilaton:
\begin{eqnarray}
\label{esv1}
{\textstyle{1\over 2}}
\, (d-1) \, \dot H &=&\hskip-3pt -{\textstyle{1\over 2}}\,\dot \phi^2 \,, \\[0.7ex]
\label{esv2}
{\textstyle{1\over 2}} \,d\, (d-1) \, H^2  &=&\hskip-3pt\phantom{-}
 {\textstyle{1\over 2}} \,\dot \phi^2 + V(\phi) \,, \\[1.1ex]
\label{esv3}
\ddot \phi + d\, H \, \dot \phi  + V'(\phi) &=& \,\,0 \,.
\end{eqnarray}
Note that
$\dot H \sim - \dot \phi^2 <0$, which means decelerating expansion or
accelerating contraction. On the other hand, the analogous equation
in the presence of a dilaton, (\ref{dotHeqn}), allows
the possibility that $\dot H$ vanishes. Equation (\ref{esv2}) is
analogous to (\ref{nodotHeqner}). Comparison of (\ref{esv3})
with (\ref{tachroll99}) confirms that the rolling scalar is only
affected by the addition of the dilaton-induced anti-friction ($\dot\Phi >0$).

The Einstein metric $g_{\mu\nu}^E$ is determined by the string metric
and the dilaton: $g_{\mu\nu}^E = \exp (-{4\over d-1}\Phi) \, g_{\mu\nu}$.
For a fixed string metric, the Einstein metric goes to zero if the dilaton
expectation value goes to infinity. This corresponds to infinite string coupling.

\section{Tachyon-driven rolling and the string metric}

We now consider a general class of potentials $V(T)$ for a tachyon $T$
that satisfy the condition $V(0)=0$ and can be written as
\begin{equation}
\label{thepottotuse}
V(T) = - {\textstyle{1\over 2}} m^2 T^2 + \mathcal{O}(T^3)\,.
\end{equation}
We build a solution
where $T\to 0$ for $t\to -\infty$, and the field rolls
to positive values:
\begin{equation}
\label{tansdkl}
T(t) = e^{mt} + \sum_{n\geq 2} t_n e^{n\,mt} \,.
\end{equation}
The first term in this ansatz is the solution to the linearized
tachyon equation of motion. The arbitrary constant multiplying this term can be
absorbed, as we did, by a redefinition of time. The exponentials
in the sum are subleading to $e^{mt}$ for large negative $t$.
We say that the tachyon drives the rolling if the other fields, in this
case $H(t)$ and $\Phi(t)$, have solutions with exponentials
subleading to $e^{mt}$:
\begin{equation}
\label{dflkvfkjvb}
\Phi(t) = \sum_{n\geq 2}  \phi_n \, e^{n\,mt} \,, \qquad
H(t) = \sum_{n\geq 2}  h_n \, e^{n\,mt} \,.
\end{equation}
Given (\ref{tansdkl}), the dilaton equation
(\ref{dilroll99}) gives
\begin{equation}
\label{leading_dilaton}
\Phi(t) = {\textstyle{1\over 8}} \, e^{2mt} + \mathcal{O}(e^{3mt}) \,.
\end{equation}
This leading behavior is valid for
all potentials of the form (\ref{thepottotuse}).  The dilaton
begins to run towards {\em stronger} coupling.
Evaluating the right-hand side of equation (\ref{dotHeqn}) we see
that
\begin{equation}
-\dot T^2 + 2\, \ddot \Phi  =  -m^2 \, e^{2mt}
+ 2 \cdot {\textstyle{1\over 8}} \cdot (4m^2)
e^{2mt} + \mathcal{O}(e^{3mt}) = 0 \cdot \, e^{2mt}+ \mathcal{O}(e^{3mt})\,.
\end{equation}
Since the other term on the right-hand side,
$H\dot\Phi \sim e^{4mt}$, we deduce that $\dot H \sim e^{3mt}$
and therefore the contribution of order $e^{2mt}$ to $H$ vanishes: $h_2 =0$.
The string metric is not affected to this order.
This is actually the beginning of a pattern:
we now prove that $H(t)$ vanishes identically
for tachyon-induced rolling.
Adding equations (\ref{dotHeqn}) and (\ref{nodotHeqner}) we find
\begin{equation}
\dot H = - (\,d\,H - 2\dot \Phi) \, H \,.
\end{equation}
Now assume that $h_2= h_3= \cdots = h_N=0$ for some $N\geq 2$.
Since $\dot\Phi \sim  e^{2mt}$, the above equation gives
$\dot H \sim e^{(N+3) \,mt}$,
which implies that $h_{N+1} =0$. By induction, $H(t)$
vanishes identically.

We now reconsider the equations of motion with
$H=0$.  The gravitational equations (\ref{dotHeqn}) and (\ref{nodotHeqner})
give a single equation, $\ddot \Phi = {1\over 2}\,\dot T^2$. Additionally,
we have the equations of motion (\ref {tachroll99}) and (\ref{dilroll99}).
With small rearrangements, the equations are:
\begin{eqnarray}
\label{feqn}
\ddot \Phi &=& {\textstyle{1\over 2}}\,  \dot T^2 \,, \\
\label{thrxxeqn}
  2\,\dot \Phi^2   &=& {\textstyle{1\over 2}}\,  \dot T^2+ V(T)\,,\\[0.1ex]
\label{seqxxn}
\ddot T - 2\,\dot \Phi  \dot T  + V'(T) &=& 0\,.
\end{eqnarray}
Since $\ddot\Phi \geq 0$, the dilaton velocity $\dot \Phi (t)$
never decreases. Given that   $\dot \Phi (t) >0$ for sufficiently
early times (see (\ref{leading_dilaton})), the dilaton $\Phi(t)$
increases without bound.  If the evolution is regular, $\Phi\to
\infty $ as $t\to \infty$ (the universe takes infinite time to
crunch).  More generally, the evolution produces a singular point
at some finite time for which, as we shall see, both $\dot \Phi$
and $\Phi$ become infinite.  Note also the complete correspondance
between the above equations and equations (\ref{esv1}),
(\ref{esv2}), and (\ref{esv3}) for an ordinary scalar coupled to
gravity. The sets of equations match, up to constants, when we set
$H\sim -\dot\Phi$.  Out of the three equations above, the last two
suffice. Taking the time derivative of (\ref{thrxxeqn}) and using
(\ref{seqxxn}), we find that (\ref{feqn}) holds as long as
$\dot\Phi \not=0$.   
The rolling of ordinary scalars with negative potentials was studied 
by Felder{\em ~et.al.}\cite{Felder:2002jk}, who noted that 
the final state is roughly independent of the shape of the potential. 
Given the correspondance 
with dilaton/tachyon rolling, this is also true in our problem.

We derived the final equations (\ref{seqxxn}) and (\ref{thrxxeqn})
using a class of initial conditions that implied $H=0$.
These equations, viewed
as the original equations with the {\em ansatz} $H=0$, allow more
general initial conditions.  For an initial time $t_i$
we can take arbitrary $T(t_i)$
and $\dot T(t_i)$ as long as
\begin{equation}
(\dot T^2 +  2V(T))\bigl|_{t_i}\, \geq 0\,.
\end{equation}
The evolution is fixed by choosing a square-root
branch for
$\dot\Phi$ in (\ref{thrxxeqn}). Since
$\dot\Phi$
is positive for tachyon-driven rolling,  we take
\begin{equation}
\label{hgke}
2\dot\Phi
=\,\sqrt{\dot T^2 + 2 V(T)} \,.
\end{equation}
This enables us to rewrite (\ref{seqxxn}) as a second-order nonlinear
differential equation for the tachyon alone, an equation that is quite
convenient for numerical integration:
\begin{equation}
\label{sodi}
\ddot T - \sqrt{\dot T^2 + 2 V(T)} \,\, \dot T + V'(T) = 0 \,.
\end{equation}

The general rolling problem can be reduced to the
problem of solving a first-order
nonlinear differential equation.  For this we consider the ``energy" $E$
defined as
\begin{equation}
\label{csiagf}
E \equiv \mathcal{E}^2  \equiv {\textstyle{1\over 2}} \dot T^2 +  V(T)  \,.
\end{equation}
One readily checks that
\begin{equation}
\label{derenx}
{d E\over dt} = \dot T\dot T (2\dot\Phi) \, \,.
\end{equation}
Since $\dot\Phi >0$, $E$ can only
increase.
The desired equation
arises by rewriting (\ref{derenx}) as
\begin{equation}
\label{deren}
{d E\over dt} =   \pm\sqrt{E-V}\,\,  {dT\over dt}\, 2\sqrt{E}\, \quad
\to \quad {d E\over dT} =  \pm 2 \sqrt{E(E-V)} \, \,.
\end{equation}
This is an equation for $E(T)$.
The sign choice arises from solving for $\dot T$ in terms of $E$ and $V$.
During evolution the sign must be changed
each time $\dot T$ goes through zero. We will use the above mostly when $\dot T>0$,
so we will take the plus sign. The   equation  becomes a little
simpler in terms of $\mathcal{E} = \sqrt{E}$:
\begin{equation}
\label{derendf}
{d \mathcal{E}\over dT} =   \sqrt{\mathcal{E}^2-V} \, \,.
\end{equation}
Equipped with $\mathcal{E}(T)$, one finds
$T(t)$ by solving the first-order linear equation that follows from (\ref{csiagf}).

A reverse engineering problem can also be solved. Suppose we are given
a tachyon rolling solution specified by a function $T(t)$ that has an inverse
$t(T)$. It is then possible to find the associated dilaton
$\Phi(t)$ and the potential $V(T)$. We use (\ref{feqn}) to find $\dot\Phi(t)$
by integration, and (\ref{thrxxeqn}) to find $V(t)$, which gives
the potential $V(t(T))$. As a simple illustration we take the leading solution
in (\ref{tansdkl}) to be exact: $T(t)= e^{mt}$.  Setting integration constants
to zero we find
\begin{equation}
T(t) = e^{mt}\,, \quad \Phi(t) = {1\over 8} e^{2mt}\,, \quad  V(T)= m^2 \Bigl(
-{1\over 2} T^2 + {1\over 8}T^4\Bigr) \,.
\end{equation}
In this solution the crunch happens at infinite string time (but finite Einstein
time). Related rolling solutions have been considered using two-dimensional
Liouville field theory to provide conformal invariant
sigma model with spacetime background
fields that typically include a linear dilaton and a constant string
metric~\cite{Tseytlin:1990mz,Strominger:2003fn,Kluson:2003xn,DaCunha:2003fm}.
In some of these solutions
$T(t) = e^{mt}$ and the linear
dilaton vanishes. This is unexpected given our analysis, which shows that
the dilaton is sourced.  It would be interesting to use this discrepancy to
find constraints on the form of the effective action for the coupled system of
fields.

\section{Finite-time crunch with negative scalar potentials}

We now show
that for non-positive potentials $V(T) \leq 0$, if $\dot \Phi (t_0) >0$ for
some time $t_0$ then $\dot \Phi (t_*) =\infty$ for some finite
time $t_* >t_0$.  To do this we
combine (\ref{feqn}) and (\ref{thrxxeqn}) to write
\begin{equation}
\label{threqndf}
-{\ddot \Phi \over {~\dot \Phi}^2} =  - 2  + {V(T) \over{\dot \Phi}^2} \,.
\end{equation}
Integrating both sides of the equation
 from an initial time $t_0$ up to a time $t$ we find
\begin{equation}
\label{threqndf}
{1 \over \dot \Phi (t)} = {1 \over \dot \Phi (t_0)} - 2 \,(t-t_0)   +\int_{t_0}^t
~ {V(T(t')) \over{\dot \Phi}^2(t')}\, dt' \,.
\end{equation}
To have a divergent $\dot \Phi$ we need the terms on the right-hand
side to add up to zero.
If we ignore the integral on the right-hand side, the first two terms
cancel for $t=t_1$, with  $t_1-t_0= 1/(2\dot \Phi (t_0))$.  Since the integral
vanishes at $t=t_0$ and can only decrease afterwards, the cancellation will actually
occur for a time earlier than $t_1$:
\begin{equation}
t_* \leq t_0 + {1 \over  2\dot\Phi (t_0)}\,.
\end{equation}
 This is what we wanted to prove.  It follows from (\ref{hgke}) and $V\leq 0$
that as $\dot \Phi \to \infty$ we also have $\dot T \to \pm \infty$.  The
time evolution reaches a singular point at finite time.

To understand how the dilaton $\Phi$ itself  diverges we can do
an estimate that proves to be self-consistent.  The integral term
in (\ref{threqndf}) is assumed to be negligible.  This is certainly
the case if $V(T)$ is also bounded below since then, the integral is negligible
for sufficiently large $\dot\Phi$.  In fact, we will see
that the integral is negligible under far more general circumstances.
It then follows that
\begin{equation}
\dot \Phi (t) \simeq  {1 \over 2(t_* - t)}\,\quad \hbox{and}
\quad \dot T (t) \simeq  {\pm 1 \over (t_* - t)}\,.
\end{equation}
For such solutions the tachyon and dilaton diverge logarithmically:
\begin{equation}
\Phi(t) = -{\textstyle{1\over 2}}\ln\, (t_*  - t) + \Phi_0
\, \qquad T(t) = \mp \ln\, (t_* - t) + T_0\, .
\end{equation}
For any polynomial potential $V(T)$ the integrand
in (\ref{threqndf}) is of the form $(t_*-t)^2 V(\ln (t_* - t))$ and goes to zero
as we approach collapse, thus justifying our approximations.
Note that
$\dot T ^2$ is much larger than both $V(T)$ and $V'(T)$.
This fact alone implies finite time collapse: the tachyon
equation (\ref{sodi}) becomes $\ddot T \mp \dot T^2  \simeq 0$,
whose general solution describes a  $\dot T$ that diverges at
an adjustable finite time.
Note that the dilaton prefactor in the spacetime action,
$e^{-2\Phi} \to  e^{-2\Phi_0} \cdot (t_*  - t) \,$,
vanishes linearly with time as we approach the collapse.
The string coupling becomes infinity, the Einstein
metric crunches, and the value of the on-shell action (\ref{sigma_action_os})
goes to zero.  Since the Einstein metric becomes much smaller than the string metric
as the dilaton diverges, the collapse also occurs in finite time in Einstein frame.

\sectiono{Crunching with arbitrary potentials}

In this section we consider rolling solutions for
rather general potentials $V(T)$, not necessarily
tachyonic. As before, we take
$H=0$ and $\dot \Phi > 0$; these conditions
ensure that we are dealing with a problem
qualitatively related
to tachyon-induced rolling. As a warmup we consider
the case where the potential is positive and bounded
and show that for a sufficiently large initial tachyon
velocity the Einstein metric crunches in finite
time.  We then discuss a related question for more
general potentials: is there an initial tachyon velocity
$\dot T >0$ for which  $T=\infty$
(and crunching) is reached in finite
time even if $V(T\to\infty)\to \infty$?
We discuss a set of tools that enable one to
approach this question systematically, at least in a case by
case basis. We find that potentials
of the form $V(T) \sim \exp (nT)$ with $n\geq 2$ are too
steep, and no positive tachyon velocity allows the tachyon
to reach $T=\infty$.
%For potentials diverge slower than $e^{2
%T}$, we find that the behaviors of the fields are totally
%determined by the initial velocity. Analytical results are given
%for bounded potentials. However, for the potentials diverge as
%$e^{2 T}$ or even faster, the anti-friction force is suppressed by
%the potential at large $T$. Therefore, the tachyon field always
%turns back at some finite time regardless the initial conditions.
%It finally hits negative singularity at some finite time.
We also analyze in detail the simple potential
$V=-\frac{1}{2} T^2+\frac{1}{8} T^4$.

\medskip
\noindent
\underbar{Crunching with bounded potentials}.
We claim that for a bounded potential $0\leq V(T) < \beta^2$
with bounded derivative $V'(T)< \gamma^2$, there is an initial
tachyon velocity $\dot T(t=0)$ for which crunching occurs in finite time.
Here is a short proof. Take $\dot
T(0)=\sqrt{\alpha^2+2\beta^2}$ with $\alpha^2>\gamma^2$. Since
the energy $E$ (see (\ref{derenx})) cannot decrease in time, for $t>0$:
\begin{equation}
2 E(t)= \dot T ^2(t)+ 2 V(T(t))>  2 E(0)= \alpha^2+2\beta^2 +2
V(T(0)) \geq \alpha^2+2\beta^2\,.
\end{equation}
Therefore,
$\dot T ^2(t)> \alpha^2+2\beta^2- 2 V(T(t)) >\alpha^2$
and, as a result, $\dot T ^2(t)> \alpha^2$
for all times. The tachyon equation of motion (\ref{sodi}) then gives
\begin{equation}
\ddot T(t)> \sqrt{\alpha^2 +2 V(T)}\, \alpha-\gamma^2 \geq
\alpha^2-\gamma^2>0.
\end{equation}
Since $\ddot \Phi=\frac{1}{2} \dot T^2$, one finds
$\dddot\Phi(t)= \dot T(t)\, \ddot T(t)>0$, so $\dot
\Phi(t)$ is convex and grows without bound. On the other hand,
with our bounds, equation (\ref{threqndf}) gives
\begin{equation}
\label{pdv}
{1 \over \dot \Phi (t)}
%= {1 \over \dot \Phi (t_0)} -
%2 \,(t-t_0) +\int_{t_0}^t ~ {V(T(t')) \over{\dot \Phi}^2(t')}\,
%dt'
< {1 \over \dot \Phi (t_0)} -
\Big(2-\frac{\beta^2}{\dot\Phi^2(t_0)}\Big) \,(t-t_0)\,.
\end{equation}
Since $\dot \Phi$ grows without bound, there is a time $t_0$ for which
$2-\frac{\beta^2}{\dot\Phi^2(t_0)}>0$. Then at some time $t_1$,
the right hand side of equation (\ref{pdv}) vanishes. Therefore,
at some finite time $t_*<t_1$, $\dot \Phi (t_*)= \infty$.

\medskip
\noindent
\underbar{General techniques}.
We now consider a general class of  potentials
$V(T)$, well-defined for
all $T$, and unbounded above as $T$ grows positive and large.
We examine an initial configuration with some fixed value
$T(t_0)$ and variable initial velocity $\dot T (t_0) >0$.
We wish to find out if the tachyon reaches $T=\infty$ and if
it does so in finite time, causing the dilaton to diverge and
the Einstein metric to crunch.
We find that, typically,
there is a critical tachyon velocity for which it takes
infinite time to reach $T=\infty$. For velocities larger than
critical, $T=\infty$ is reached in finite time. For velocities
smaller than critical the tachyon evolution gives a turning point.

%There are some tools that help analyze our differential
%equation $\frac{d\mathcal E}{dT}=\sqrt{\mathcal
%E^2-V(T)}\equiv h(T,\mathcal E)$.
Three curves can be defined
in the $(T, \mathcal{E})$ plane and help us understand
the integral curves $\mathcal{E}(T)$  that solve our first-order
differentail equation $\frac{d\mathcal E}{dT}=\sqrt{\mathcal
E^2-V(T)}\equiv h(T,\mathcal E)$:
\begin{itemize}

\item $h( T,\mathcal E)=0$ is the \emph{turning point} curve.
%The integral curves lie above this curve. If an integral curve
%reaches the turning point curve, $\dot T=0$, and we have a turning point.
%, $\mathcal E(T)$ is increasing; below this curve,
%$\mathcal E(T)$ is decreasing.

\item $\frac{d}{dT} h(T,\mathcal E)=0$ is the \emph{inflection}
curve. It separates a region where the integral curves are convex
from a region where they are concave.

\item $\mathcal E^2-V(T)=f(T)$, with $f$ specified below, is the
\emph{separating} curve. Any integral curve starting above the
separating curve will remain above it.
\end{itemize}

Since $\frac{d\mathcal E}{dT}=\sqrt{\mathcal
E^2-V}=\frac{1}{\sqrt 2} |\dot T(t)|$ the turning point curve
is the locus of points where we get  turning points for the tachyon
time evolution. If an integral curve hits the turning point curve,
the time evolution of the tachyon has a turning point.
  Moreover, since $\frac{d}{dT} \dot
T=\frac{\ddot T}{\dot T}$,  the inflection curve also controls the
convexity or concavity of $T(t)$.

Consider the curve
$\mathcal E^2-V(T)=f(T)$. On this curve the slope of the
integral curve is $\sqrt{f(T)}$. Moreover, the slope of this
curve itself is $\frac{f'+V'}{2\sqrt{f+V}}$. In order to be
a separating curve we require the former to be larger than the latter:
\begin{equation}
\sqrt{f(T)}\geq \frac{f'+V'}{2\sqrt{f+V}}\,\,.
\end{equation}
The equality gives an integral curve -- integral curves separate because
they cannot cross, but are hard to find.
If $V(T)\geq 0$, a suitable
$f(T)$ is obtained by setting:
\begin{equation}
\label{fdef} \sqrt{f(T)}= \frac{f'+V'}{2\sqrt{f}}\quad \to \quad  2
f-f'=V'.
\end{equation}
For a polynomial $V(T)$, a convenient
choice is
$f(T)=\sum_{n=1} {1\over 2^n}\,{d^nV \over dT^n}$.

\begin{figure}
\centerline{\hbox{\epsfig{figure=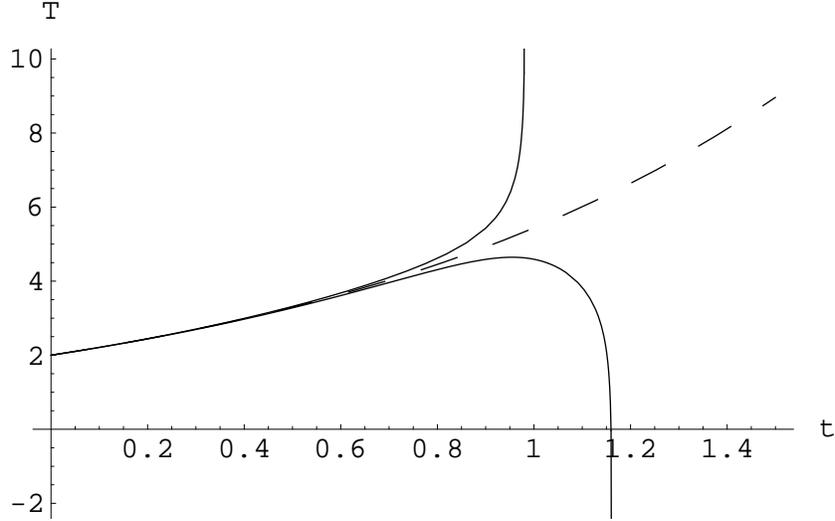, height=7cm}}}
\caption{Tachyon histories $T(t)$ in the potential $V(T)=-\frac{1}{2}T^2+\frac{1}{8}
T^4$, starting with $T(t=0)=2$. The
critical trajectory is dashed, begins with
critical velocity $\dot T(0)=2$, and crunches
at infinite time. The solid line turning upwards
starts with $\dot T(0)=2.01$ and hits $T=\infty$ at $t\simeq 0.9808$. The
solid line turning downward starts with initial velocity $\dot T(0)=1.99$,
encounters a turning point, and actually reaches minus infinity at
$t\simeq 1.1609$. } \label{diffvel}
\end{figure}

\medskip
\noindent
\underbar{A worked out example}.
We illustrate the  above discussion with the potential
$V=-\frac{1}{2} T^2+\frac{1}{8} T^4$, which is tachyonic near $T=0$,
vanishes for $T=\pm 2$, and grows arbitrarily large for large $|T|$.
We consider arbitrary velocities for a tachyon for which $T=2$ for $t=0$.
The potential was chosen so that it has an easily obtained
solution with critical velocity: $T(t)=2 \exp(t)$. This is, in fact,
the tachyon-induced rolling solution that starts at $T=0$ for $t=-\infty$.
Here $T(0)=2$, $\dot T (0)= 2$,
  $\mathcal{E}(T=2)=
\sqrt{2}$, and the solution reaches $T=\infty$ at $t=\infty$.
We have verified numerically that solutions
with larger initial velocity reach $T=\infty$ in finite time, while
solutions with lesser initial velocity encounter a turning point.
In Fig.~\ref{diffvel}, we show the critical trajectory $T(t)$ and two
additional solutions corresponding to initial velocities slightly
higher and slightly lower than critical.

\begin{figure}
\centerline{\hbox{\epsfig{figure=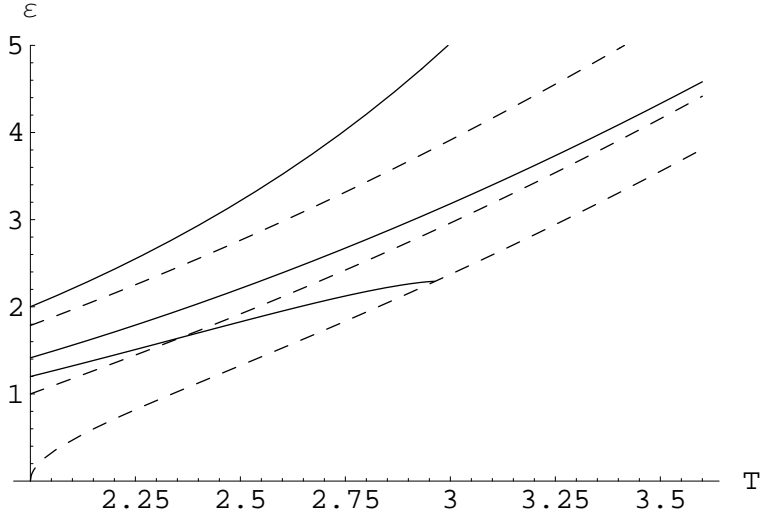, height=7cm}}}
\caption{Curves and solutions for
${d \mathcal{E}\over dT} =   \sqrt{\mathcal{E}^2-V} $ with
potential $V=-\frac{1}{2} T^2+\frac{1}{8} T^4$, plotted in
the $(T, \mathcal{E})$ plane with vertical axis at $T=2$.
From bottom to top, the three dashed curves are: the turning point
curve ($\mathcal E(T=2)=0$), the inflection curve
($\mathcal E(2)=1$), and the separating curve
($\mathcal E(2)=\sqrt{51}/4
\simeq 1.7854$).
The three solid curves are solutions.  From bottom to
top they are: a solution with subcritical $\mathcal E(2) =1.20$
that hits the turning point curve, the critical solution,
with $\mathcal E(2)=\sqrt 2$, and a solution
above the separating curve, with $\mathcal E(2) =2.0$.}
\label{charcurves}
\end{figure}

For the potential in question, the turning point curve $\mathcal E(T)= \sqrt {V(T)}$
lies below
 the inflection curve $\mathcal
E(T)=(\frac{V+\sqrt{V^2+V'^2}}{2})^{1/2}$, which in turn lies below
the separating curve ($T\geq 2$) defined with
$f(T)=-\frac{1}{16}(1+2 T-6 T^2-4 T^3)$. In Fig.~\ref{charcurves}
we plot these three curves along with three solutions. The solution
with lowest initial energy has velocity smaller than critical:
it crosses the inflection curve and  hits the turning point curve.
We also show the critical solution and a solution with velocity
larger than critical that lies above the separating curve.  By construction
any solution above the separating curve cannot have a turning point.

\medskip
\noindent
\underbar{Potentials of the form $V(T)=\exp(n T)$}.
We now show that for $V(T)=\exp(n T)$,
with $n\geq 2$, there is no initial (positive) tachyon
velocity for which the tachyon can reach $T=\infty$.
When $V(T)=\exp(n T)$ the differential equation (\ref{derendf})
is solvable. Setting $\mathcal E= \sqrt V g(T)$, we find
\begin{equation}
\label{funcg}
g'+\frac{V'}{2 V} \,g=\sqrt{g^2-1}\quad \to \quad
g'=\sqrt{g^2-1}-\frac{n}{2}\, g\,.
\end{equation}
We have a turning point at $T_*$ if $g(T_*)=1$. Dividing
both sides of the equation by $g$ and integrating,
\begin{equation}
\ln g(T)=\ln g(T_0)-\frac{n}{2} (T-T_0)+ \int_{T_0}^T dT'
%\sqrt{1-\frac{1}{g^2(T')}}
\Bigl(1-\frac{1}{g^2(T')}\Bigr)^{1/2}
< \ln g(T_0)-(\frac{n}{2}-1)
(T-T_0).
\end{equation}
For $n>2$, the right hand side vanishes at some $T_1>T_0$.
There is therefore some $T_*<T_1$  with $g(T_*)=1$, and thus a turning
point, as we wanted to show. The above argument in fact applies for
any potential $V$ such that ${V'\over V} >2$ for sufficiently large $T$.
For $n=2$, the solution of (\ref{funcg}) is
\begin{equation}
T=T_0+\frac{1}{2}\big( F(g)- F(g_0)), \hspace{5mm} F(g)\equiv\ln
(g+\sqrt{g^2-1})- g(g+ \sqrt {g^2-1}).
\end{equation}
It is readily checked  that $F(1)=-1$ and $F(g)$ decreases monotonically for
 $g>1$. Therefore, $g(T^*)=1$ for
$T^*=T_0-\frac{1}{2}\big( 1+ F(g_0))> T_0$.  This proves that all solutions
have a turning point.

%Since beyond the turning point, $\dot T(t)<0$, the governing
%differential equation becomes
%\begin{equation}
%\frac{d\mathcal E}{dT}=-\sqrt{\mathcal E^2-V}\Rightarrow
%g'(T)=-\sqrt{g^2(T)-1}-\frac{n}{2}\, g(T).
%\end{equation}
%It is easy to see that if $g(T)>0$, $g'(T)<0$ and $g''(T)>0$. On
%the other hand, $T^*$ now behaves as the initial point and
%$g(T^*)=1$ as well as $\dot T(t)<0$, therefore we conclude that
%$g(T)$ increases without bound while $T(t)$ moves all along to the
%negative infinity. At large $g(T)\gg 1$,
%\begin{equation}
%g'(T)\simeq -\Big(1+\frac{n}{2}\Big) g(T) \Rightarrow
%g(T)=e^{-(1+n/2) (T-T^*)}.
%\end{equation}
%Therefore, $\mathcal E=\mathcal E (T^*) e^{-(T-T^*)}$. With the
%relation $\frac{1}{\sqrt 2}\dot T=\frac{d}{dT} \mathcal E$, one
%readily obtains
%\begin{equation}
%T(t)=\ln\Big(e^{T^*}-\sqrt 2 \mathcal E (T^*) (t-t_0)\Big).
%\end{equation}
%Therefore, $T(t)$ hit minus infinity at some finite time.
%

\sectiono{Conclusions}

Our analysis of tachyon-induced rolling has revealed
two general facts: 1) the string metric is
constant and, 2) the dilaton rolls toward stronger coupling.
These facts match precisely the properties of
the candidate tachyon vacuum identified in~\cite{HZ}.
Consider fact one. In the tachyon vacuum both the tachyon and the dilaton
take expectation values, and so do an infinite number
of massive fields.  The string metric, however, is not sourced and
need not acquire an expectation value -- this is guaranteed
by rather general string field theory universality
arguments~\cite{Sen:1999xm,HZ}. The cosmological constancy
of the string metric appears
to be the sigma model version of the universality result.
Consider now fact two. It was shown
in~\cite{HZ} that the dilaton expectation value in the
candidate solution corresponds
to stronger string coupling.   The qualitative agreement makes it
plausible that the rolling solutions discussed here
represent rolling towards the tachyon vacuum conjectured in~\cite{HZ}.
In the case of open strings,  the end-product of the
rolling solution is  different but somewhat related to the tachyon
vacuum. Our rolling solutions represent an Einstein metric big crunch
or closed strings at infinite coupling.
The tachyon vacuum may represent the dissappearance of
dynamical spacetime.  We feel these two states could be
related. The crunch certainly lies beyond the applicability of the
action (\ref{sigma_action}), which should be
supplemented by terms of higher order in $\alpha'$.
The generality of the evolution and the almost complete
independence on the details of the tachyon
potential\footnote{For comments on the specific form of the
tachyon potential in closed string theory see Figure~1 and
the conclusion section of~\cite{HZ}.} %bz
 suggest to us
that the cosmological solutions presented here are relevant, modulo
some stringy resolution of the big crunch singularity.  It would be
rather interesting if the stringy resolution would push the crunch
to infinite time.
A big crunch, followed by a big bang,
is the key element in cyclic universe models~\cite{Steinhardt:2004gk}.
The crunch is induced by a scalar field rolling down a {\em negative} potential
with a steep region -- the rest of the potential is largely undetermined.
Negative, initially steep potentials, are the hallmark of bulk closed
string tachyons. We found that generally a big crunch  ensues, although
in our case the gravitational part of the solution is carried by the dilaton,
and thus the crunch has the alternative interpretation of a closed string theory at
infinite coupling.  It is tempting to speculate that closed string
tachyons may play a role in cyclic universe scenarios -- the central difficulty
remaining the mysterious transition from a big crunch to a big bang. In such
studies it would be useful to focus on tachyonic heterotic models and Type-0 strings.

If the vacuum of the bulk closed string tachyon truly represents the
demise of fluctuating spacetime, understanding properly this state and
how it fits into a consistent cosmology would give invualuable insight
into the mechanisms by which a universe could come into existence.  The
tachyon vacuum would be roughly imagined to be the state of a universe
before the Big Bang.

\bigskip
\noindent
{\bf Acknowledgements}
We are indebted to Justin Khoury for many illuminating
discussions and significant guidance
on the subject of cosmological solutions. We would also
like to acknowledge useful conversations with K.~Hashimoto,
M.~Headrick, H. Liu, A.~Sen,  A. Tseytlin,
and Y.~Okawa.


\bigskip

\begin{thebibliography}{99}

\small

%\cite{Taylor:2003gn}
%\cite{Sen:2004nf}
\bibitem{reviews}
%\bibitem{Sen:2004nf}
  A.~Sen,
  ``Tachyon dynamics in open string theory,''
  arXiv:hep-th/0410103.
  %%CITATION = HEP-TH 0410103;%%
%\bibitem{Taylor:2003gn}
W.~Taylor and B.~Zwiebach, ``D-branes, tachyons, and string field
theory,'' arXiv:hep-th/0311017;
%%CITATION = HEP-TH 0311017;%%
%
%\cite{DeSmet:2001af}
%\bibitem{DeSmet:2001af}
P.~J.~De Smet, ``Tachyon condensation: Calculations in string
field theory,'' arXiv:hep-th/0109182;
%%CITATION = HEP-TH 0109182;%%
%
%\cite{Ohmori:2001am}
%\bibitem{Ohmori:2001am}
K.~Ohmori, ``A review on tachyon condensation in open string field
theories,'' arXiv:hep-th/0102085;
%%CITATION = HEP-TH 0102085;%%
%
%
%\cite{Bonora:2003xp}
%\bibitem{Bonora:2003xp}
L.~Bonora, C.~Maccaferri, D.~Mamone and M.~Salizzoni, ``Topics in
string field theory,'' arXiv:hep-th/0304270.
%%CITATION = HEP-TH 0304270;%%

\bibitem{Headrick:2004hz}
  M.~Headrick, S.~Minwalla and T.~Takayanagi,
  ``Closed string tachyon condensation: An overview,''
  Class.\ Quant.\ Grav.\  {\bf 21}, S1539 (2004)
  [arXiv:hep-th/0405064];
%%\cite{Adams:2005rb}
%\bibitem{Adams:2005rb}
  A.~Adams, X.~Liu, J.~McGreevy, A.~Saltman and E.~Silverstein,
  ``Things fall apart: Topology change from winding tachyons,''
  arXiv:hep-th/0502021.
  %%CITATION = HEP-TH 0502021;%%
  %%CITATION = HEP-TH 0405064;%%
%\bibitem{Adams:2001sv}
%A.~Adams, J.~Polchinski and E.~Silverstein,
%``Don't panic! Closed string tachyons in ALE space-times,''
%JHEP {\bf 0110}, 029 (2001)
%[arXiv:hep-th/0108075].
%%%CITATION = HEP-TH 0108075;%%
%%\cite{Gregory:2003yb}
%%\bibitem{Gregory:2003yb}
%R.~Gregory and J.~A.~Harvey,
%``Spacetime decay of cones at strong coupling,''
%Class.\ Quant.\ Grav.\  {\bf 20}, L231 (2003)
%[arXiv:hep-th/0306146].
%%%CITATION = HEP-TH 0306146;%%
%%\cite{Headrick:2003yu}
%%\bibitem{Headrick:2003yu}
%M.~Headrick,
%``Decay of C/Z(n): Exact supergravity solutions,''
%JHEP {\bf 0403}, 025 (2004);
%[arXiv:hep-th/0312213].
%%%CITATION = HEP-TH 0312213;%%
Y.~Okawa and B.~Zwiebach,
``Twisted tachyon condensation in closed string field theory,''
JHEP {\bf 0403}, 056 (2004)
[arXiv:hep-th/0403051].
%%CITATION = HEP-TH 0403051;%%
%\cite{Bergman:2004st}
%\bibitem{Bergman:2004st}
O.~Bergman and S.~S.~Razamat, ``On the CSFT approach to localized
closed string tachyons,'' arXiv:hep-th/0410046.
%%CITATION = HEP-TH 0410046;%%





\bibitem{HZ}
H.~Yang and B.~Zwiebach, ``A closed string tachyon vacuum~?",
to appear.

%\cite{Das:1986cz}
\bibitem{sdas}
  S.~R.~Das and B.~Sathiapalan,
  ``String Propagation In A Tachyon Background,''
  Phys.\ Rev.\ Lett.\  {\bf 56}, 2664 (1986).
  %%CITATION = PRLTA,56,2664;%%%\cite{Callan:1986ja}
%\bibitem{Callan:1986ja}
  C.~G.~.~Callan and Z.~Gan,
  ``Vertex Operators In Background Fields,''
  Nucl.\ Phys.\ B {\bf 272}, 647 (1986).
  %%CITATION = NUPHA,B272,647;%%
%\cite{Tseytlin:1991bu}
%\bibitem{Tseytlin:1991bu}
  A.~A.~Tseytlin,
  ``On the tachyonic terms in the string effective action,''
  Phys.\ Lett.\ B {\bf 264}, 311 (1991).
  %%CITATION = PHLTA,B264,311;%%


%\cite{Gasperini:2002bn}
\bibitem{Gasperini:2002bn}
  M.~Gasperini and G.~Veneziano,
  ``The pre-big bang scenario in string cosmology,''
  Phys.\ Rept.\  {\bf 373}, 1 (2003)
  [arXiv:hep-th/0207130].
  %%CITATION = HEP-TH 0207130;%%


%\cite{Dine:2003ca}
\bibitem{Dine:2003ca}
  M.~Dine, E.~Gorbatov, I.~R.~Klebanov and M.~Krasnitz,
  ``Closed string tachyons and their implications for non-supersymmetric
  strings,''
  JHEP {\bf 0407}, 034 (2004)
  [arXiv:hep-th/0303076].
  %%CITATION = HEP-TH 0303076;%%

%\cite{Suyama:2003as}
\bibitem{Suyama:2003as}
  T.~Suyama,
  ``On decay of bulk tachyons,''
  arXiv:hep-th/0308030.
  %%CITATION = HEP-TH 0308030;%%

%\cite{Felder:2002jk}
\bibitem{Felder:2002jk}
  G.~N.~Felder, A.~V.~Frolov, L.~Kofman and A.~V.~Linde,
  %``Cosmology with negative potentials,''
  Phys.\ Rev.\ D {\bf 66}, 023507 (2002)
  [arXiv:hep-th/0202017].
  %%CITATION = HEP-TH 0202017;%%


%\cite{Kostelecky:1992vg}
\bibitem{Kostelecky:1992vg}
  V.~A.~Kostelecky and M.~J.~Perry,
  ``Condensates and singularities in string theory,''
  Nucl.\ Phys.\ B {\bf 414}, 174 (1994)
  [arXiv:hep-th/9302120].
  %%CITATION = HEP-TH 9302120;%%

%\cite{Polchinski:1998rq}
\bibitem{Polchinski:1998rq}
  J.~Polchinski,
  ``String theory. Vol. 1: An introduction to the bosonic string,''
Cambrigde University Press (1998).

%\cite{Tseytlin:1990mz}
\bibitem{Tseytlin:1990mz}
  A.~A.~Tseytlin,
  ``On The Structure Of The Renormalization Group Beta Functions In A Class Of
  Two-Dimensional Models,''
  Phys.\ Lett.\ B {\bf 241}, 233 (1990).
  %%CITATION = PHLTA,B241,233;%%


%\cite{Strominger:2003fn}
\bibitem{Strominger:2003fn}
  A.~Strominger and T.~Takayanagi,
  ``Correlators in timelike bulk Liouville theory,''
  Adv.\ Theor.\ Math.\ Phys.\  {\bf 7}, 369 (2003)
  [arXiv:hep-th/0303221];
  %%CITATION = HEP-TH 0303221;%%
%\cite{Schomerus:2003vv}
%\bibitem{Schomerus:2003vv}
  V.~Schomerus,
  ``Rolling tachyons from Liouville theory,''
  JHEP {\bf 0311}, 043 (2003)
  [arXiv:hep-th/0306026];
  %%CITATION = HEP-TH 0306026;%%
%\cite{Kluson:2003xn}

\bibitem{Kluson:2003xn}
  J.~Kluson,
  ``The Schrodinger wave functional and closed string rolling tachyon,''
  Int.\ J.\ Mod.\ Phys.\ A {\bf 19}, 751 (2004)
  [arXiv:hep-th/0308023];
  %%CITATION = HEP-TH 0308023;%%
%\cite{Hikida:2004mp}
%\bibitem{Hikida:2004mp}
  Y.~Hikida and T.~Takayanagi,
  ``On solvable time-dependent model and rolling closed string tachyon,''
  Phys.\ Rev.\ D {\bf 70}, 126013 (2004)
  [arXiv:hep-th/0408124].
  %%CITATION = HEP-TH 0408124;%%



%\cite{DaCunha:2003fm}
\bibitem{DaCunha:2003fm}
  B.~C.~Da Cunha and E.~J.~Martinec,
  %``Closed string tachyon condensation and worldsheet inflation,''
  Phys.\ Rev.\ D {\bf 68}, 063502 (2003)
  [arXiv:hep-th/0303087].
  %%CITATION = HEP-TH 0303087;%%



\bibitem{Steinhardt:2004gk}
  P.~J.~Steinhardt and N.~Turok,
  ``The cyclic model simplified,''
  arXiv:astro-ph/0404480;
  %%CITATION = ASTRO-PH 0404480;%%
%\bibitem{Khoury:2004xi}
  J.~Khoury,
  ``A briefing on the ekpyrotic / cyclic universe,''
  arXiv:astro-ph/0401579;
  %%CITATION = ASTRO-PH 0401579;%%
%\cite{Khoury:2003rt}
%\bibitem{Khoury:2003rt}
  J.~Khoury, P.~J.~Steinhardt and N.~Turok,
  ``Designing cyclic universe models,''
  Phys.\ Rev.\ Lett.\  {\bf 92}, 031302 (2004)
  [arXiv:hep-th/0307132];
  %%CITATION = HEP-TH 0307132;%%
%\cite{Khoury:2001bz}
%\bibitem{Khoury:2001bz}
  J.~Khoury, B.~A.~Ovrut, N.~Seiberg, P.~J.~Steinhardt and N.~Turok,
  ``From big crunch to big bang,''
  Phys.\ Rev.\ D {\bf 65}, 086007 (2002)
  [arXiv:hep-th/0108187].
  %%CITATION = HEP-TH 0108187;%%




%





%\cite{Zwiebach:1992ie}
\bibitem{Zwiebach:1992ie}
B.~Zwiebach,
``Closed string field theory: Quantum action and the B-V master equation,''
Nucl.\ Phys.\ B {\bf 390}, 33 (1993)
[arXiv:hep-th/9206084].
%%CITATION = HEP-TH 9206084;%%






%\cite{Sen:1999xm}
\bibitem{Sen:1999xm}
  A.~Sen,
  ``Universality of the tachyon potential,''
  JHEP {\bf 9912}, 027 (1999)
  [arXiv:hep-th/9911116].
  %%CITATION = HEP-TH 9911116;%%





\end{thebibliography}
\end{document}